\documentclass[sigconf]{acmart}

\usepackage{acmart-taps}
\usepackage{graphicx}

\usepackage{pifont}
\newcommand{\cmark}{\ding{51}} 
\newcommand{\xmark}{\ding{55}} 

\AtBeginDocument{%
  }

\copyrightyear{2026}
\acmYear{2026}
\setcopyright{cc}
\setcctype{by}
\acmConference[CHI '26]{Proceedings of the 2026 CHI Conference on Human Factors in Computing Systems}{April 13--17, 2026}{Barcelona, Spain}
\acmBooktitle{Proceedings of the 2026 CHI Conference on Human Factors in Computing Systems (CHI '26), April 13--17, 2026, Barcelona, Spain}
\acmPrice{}
\acmDOI{10.1145/3772318.3791015}
\acmISBN{979-8-4007-2278-3/2026/04}

\begin{document}

\title[Good for the Planet, Bad for Me?]{Good for the Planet, Bad for Me? Intended and Unintended Consequences of AI Energy Consumption Disclosure}

\author{Michael Klesel}
\orcid{0000-0002-2884-1819}
\affiliation{%
  \institution{Frankfurt University of Applied Sciences}
  \city{Frankfurt}
  \country{Germany}}
\email{michael.klesel@fra-uas.de}
\affiliation{%
  \institution{hessian.AI}
  \city{Darmstadt}
  \country{Germany}}

\author{Uwe Messer}
\email{uwe.messer@unibw.de}
\orcid{0000-0003-0473-5066}
\affiliation{%
  \institution{Universität der Bundeswehr München}
  \city{München}
  \country{Germany}
}

\renewcommand{\shortauthors}{Klesel \& Messer}

\begin{abstract}
  To address the high energy consumption of artificial intelligence, energy consumption disclosure (ECD) has been proposed to steer users toward more sustainable practices, such as choosing efficient small language models (SLMs) over large language models (LLMs). This presents a performance-sustainability trade-off for users. In an experiment with 365 participants, we explore the impact of ECD and the perceptual and behavioral consequences of choosing an SLM over an LLM. Our findings reveal that ECD is a highly effective measure to nudge individuals toward a pro-environmental choice, increasing the odds of choosing an energy efficient SLM over an LLM by more than 12. Interestingly, this choice did not significantly impact subsequent behavior, as individuals who selected an SLM and those who selected an LLM demonstrated similar prompt behavior. Nevertheless, the choice created a perceptual bias. A placebo effect emerged, with individuals who selected the "eco-friendly" SLM reporting significantly lower satisfaction and perceived quality. These results highlight the double-edged nature of ECD, which holds critical implications for the design of sustainable human-computer interactions.
\end{abstract}



\maketitle

\section{Introduction}
The impressive capabilities of large language models (LLMs) come at a substantial and often hidden resource cost \cite{deVries2023, strubellEnergyPolicyConsiderations2020, luccioni2024power}. For example, training BLOOM is estimated to have incurred over 50 tonnes of CO\textsubscript{2} emissions, and the training of GPT-3 consumed 700,000 liters of freshwater \cite{Li2025}. The growing environmental footprint of artificial intelligence (AI) has not only led to extensive academic research on green AI \cite{Schwarz2020}, but has also become a topic of public discussion. Concerns range from the energy needed for a single LLM query \cite{ODonnel2025} to the documented increase in carbon emissions from major AI providers \cite{Milmo_Google_2024, kearney2025google}. Although there are various measures of the environmental effects of AI, such as CO\textsubscript{2} emissions and water footprint, energy demand is most often used to demonstrate AI's sustainability implications \cite{luccioni2024power, Li2025}.

From an AI lifecycle perspective, existing literature on energy consumption and AI has primarily focused on model training \cite{strubellEnergyPolicyConsiderations2020}. However, while understanding the resources needed for training is important, there is a growing body of evidence indicating that inference is the larger issue. Inference -- the process of using an AI model for new data -- accounts for an estimated 60\% of AI's total energy consumption \cite[p. 15]{aljbour2024}, a figure corroborated by major industry reports \cite[][p. 24]{patterson2022carbon}. Given that even a medium-length GPT-3 response requires half a liter of water \cite{Li2025} and $\approx$ 0.5 W to generate an image \cite{luccioni2024light}, there is a pressing need to identify strategies that encourage more efficient use of these powerful models at the point of interaction. Currently, new technological concepts have been proposed to identify the most efficient model based on a query and to apply model routing \cite{Google2025ModelRouting}. However, these concepts do not consider individual autonomy and fail to raise awareness of the underlying energy demands. 

We argue that focusing on inference presents new opportunities to conduct research from a sustainable Human-Computer-Interaction (HCI) perspective that complements technical efforts \citep{laurell2025exploring, disalvo2010mapping, mankoff2007environmental}. Conceptually, prior work has used the concept of pro-environmental behavior (PEB) \citep{Hines1987, Bamberg2007}, defined as \textit{"behaviour that harms the environment as little as possible, or even benefits the environment"} \cite[][p. 309]{Steg2009}. In the context of AI use, a critical behavior that reflects PEB is the choice between alternative models (e.g., choosing GPT-4o-mini over GPT-4o). Conceptually, these models can be distinguished as energy-efficient small language models (SLMs) \cite{vannguyen2024surveysmalllanguagemodels, Jeanquartier2026} and conventional LLMs. In general, using an SLM instead of an LLM for the same task will require fewer resources and is thus more sustainable. Therefore, we consider the decision to choose a less resource-intensive SLM as a specific instance of PEB in AI use.

Although users are commonly able to choose between models\footnote{For example, an OpenAI user can choose between ChatGPT Instant, GPT Thinking, and GPT Pro. Similarly, a Google user can choose between Gemini Flash, Gemini Flash-Light, and Gemini Pro.}, the lack of transparency regarding the energy demand of these models makes it nearly impossible to make an informed choice when pursuing an energy-efficient option. Because most end-users use cloud-based services, which are often perceived as having "zero costs" \cite{siy2025litter}, this problem is exacerbated. For this reason, \citet{luccioni2024light} argue that transparency is a necessary prerequisite to raise awareness of energy demands and to enable informed choices, which can ultimately lead to a behavioral change. 

To address such transparency deficits, research and industry in other domains have implemented (mandatory) product labels that provide information on potential adverse environmental effects. For example, in 1994, the European Union introduced an energy label \cite{EU2009_125, EU2017_1369}, which uses a traffic light system for energy consumption disclosure (ECD) to effectively communicate efficiency classes of appliances to end users. Empirical studies have shown that the disclosure of environmental information helps educate consumers and influence their purchasing decisions \cite{davis2016does, ikonen2020consumer}. This disclosure paradigm has also been extended to the food sector in the form of front-of-package labels, which are designed to quickly convey information on nutritional value, quality, or sustainability \cite{muzzioli2023communicate}. Digital settings also use labeling to guide behavior, as seen with privacy labels for apps \cite{li2022understanding} and content descriptors for video games \cite{xiao2023shopping}. 

The effectiveness of ECD in other domains \cite{davis2016does, ikonen2020consumer} suggests that greater transparency could also benefit users seeking to make informed decisions about AI. Beyond individual empowerment, ECD can also address growing regulatory demands for transparency from international organizations and governments. For example, the OECD recommends the implementation of standardized energy measures for the entire AI lifecycle \cite{oecd2022measuring}, and ongoing regulatory efforts such as the EU AI Act mandate the documentation of energy consumption requirements and indicators during model training and use \cite[especially Art. 40(2) \& Art. 95(2)]{EUAIAct2024}. 

While ECD is a necessary prerequisite for PEB in model choice, choosing between an SLM and an LLM presents a direct trade-off for users because the objective performance of an SLM is typically lower than that of an LLM. This decrease in performance can be substantial. For example, GPT-3 with 2.7 billion parameters (SLM) has an average performance of 25.9 on the Massive Multitask Language Understanding (MMLU) benchmark, whereas GPT-3 with 175 billion parameters (LLM) has an average performance of 43.9 \cite{hendryckstest2021}. This reflects a performance drop of $\approx$18\% on average and presents a fundamental sociotechnical tension: the choice between environmental sustainability and optimal performance is shifted directly onto the user, who can choose for a pro-environmental option at the cost of performance. This creates a critical design challenge that this paper aims to address.

To investigate this challenge, we designed and evaluated a choice architecture that presents users with two options: (1) a high-perfor\-mance LLM and (2) an energy-efficient SLM. Our approach is grounded in multiple theoretical frameworks. First, we draw upon nudge theory \cite{ThalerSunstein2009} to examine how ECD influences initial model choice. To understand potential variations in this effect, we also consider users' pro-environmental attitude (PEA) \cite{Shen2024} as a moderating factor. Second, we examine the post-choice consequences. We use moral licensing theory \cite{blanken2015meta} to investigate whether the initial sustainable choice affects subsequent usage behavior, and concepts from placebo effect research \cite{kosch2023placebo, vaccaro2018illusion, denisova2015placebo} to explore whether the choice influences users' perceptions of the AI model's quality and performance.

Based on the empirical results of our experiment, this study makes three contributions: First, we empirically test the proposition about the effectiveness of ECD \cite{luccioni2024light}. Most notably, we show that -- contrary to the generally modest effects of nudging \cite{Szaszi2025} -- nudging individuals with ECD can have a strong effect on model choice (odds ratio > 12). Second, we demonstrate that the design of the choice interface (the ECD nudge) exerts a stronger effect on user behavior than pre-existing pro-environmental attitudes, though a small interaction effect is present. This finding underscores the power of interface design in shaping sustainable behavior. Finally, we reveal a critical trade-off, showing that while ECD promotes PEB, it can also negatively impact user satisfaction and perceived quality of the language model. Our work thus highlights a central challenge for sustainable HCI: while interface design can strongly encourage pro-environmental choices, it may do so at the expense of the user experience we aim to improve.

\section{Related Work}
\subsection{The Resource Cost of Large-Scale AI}
The performance gains of modern AI models are coupled with significant environmental and resource costs \cite{strubellEnergyPolicyConsiderations2020, luccioni2024power,jegham2025hungry, Li2025,Dauner2025}. Early investigations into this AI footprint focused on the energy consumption required for deep learning in natural language processing, revealing that the carbon emissions from training a single model could rival the entire lifecycle emissions of several cars \cite{strubellEnergyPolicyConsiderations2020}. This initial focus on \textit{build-time} or training costs has been extensively explored, with subsequent research conducting comprehensive lifecycle assessments of foundation models to account for everything from hardware manufacturing to the energy used during training \cite{Luccioni2023BLOOM}. Beyond energy, studies have also highlighted the immense water footprint of AI, primarily due to the freshwater required to cool the data centers that train models like GPT-3 \cite{Li2025, jegham2025hungry}. This body of work established a clear link between AI development and environmental impact, prompting calls for transparent tracking mechanisms to better monitor this footprint \cite{jegham2025hungry}.

While the costs of model training are substantial, a growing body of research argues that inference -- i.e., the computational cost of a model's operation -- constitutes the larger portion of AI's long-term energy footprint \cite{Samsi2023, Wilhelm2025}. This shift in focus is driven by two primary factors. First, the widespread consumer and enterprise adoption of generative AI has led to an exponential increase in the number of queries processed daily. Even if a single query consumes a small amount of energy (between $\approx 5 \times 10^{-4}$\,W and $\approx 5 \times 10^{-1}$\,W \cite{luccioni2024light}), the cumulative effect of billions of daily inferences requires a massive, energy-intensive infrastructure \cite{patterson2022carbon, aljbour2024}. Second, state-of-the-art prompting techniques designed to improve model performance often do so by increasing the computational complexity at inference time. For instance, methods such as Chain-of-Thought (CoT) prompting elicit more accurate responses by requiring a model to generate intermediate reasoning steps, thereby increasing the length of the output and the overall energy required for a single task \cite{Wei2022, Sardana2024}. This shifts the primary locus of energy consumption from a one-time training event to a continuous, user-driven operational cost, making the study of user behavior at inference time a critical area for HCI research.

A key determinant of a model's energy consumption at inference is its size, typically measured by the number of parameters \cite{jegham2025hungry, Dauner2025}. LLMs are defined by their massive parameter counts, which have grown exponentially from models like the original GPT-1 (117 million parameters) to contemporary models estimated to have over a trillion parameters \cite{Chang2024}. In response to these scaling challenges, significant research has focused on developing smaller, more efficient models. These SLMs are designed with efficiency as a primary goal, often utilizing techniques like knowledge distillation to inherit capabilities from a larger "teacher" model while maintaining a much smaller parameter count \cite{vannguyen2024surveysmalllanguagemodels}. A classic example is DistilBERT, a distilled and more efficient version of the original BERT model \cite{Sanh2020}. This creates a direct, yet complex, relationship between a model's size, performance, and energy efficiency. Therefore, for the purposes of this study, we follow prior work \cite{jegham2025hungry, Dauner2025} and use the distinction between high-parameter LLMs and smaller, more efficient SLMs as a practical proxy for high- and low-energy consumption options, respectively. 

\subsection{Supporting User Decisions Through Energy Disclosure}
Sustainable HCI has emerged as an important stream of research to provide environmental information in the interaction with technology \cite{laurell2025exploring, disalvo2010mapping}. This is particularly important when considering technologies used in everyday situations -- such as generative AI -- as it has the potential to drive behavior toward greater sustainability. Supporting and guiding users' decision-making processes aligns with the "sustainability through design" approach \cite[][p. 2123]{mankoff2007environmental}. Helping individuals make more informed decisions can help to achieve near-term sustainability gains \citep{laurell2025exploring}. Our research is situated within this tradition of leveraging design to empower sustainable decision-making in the context of AI use.

In a recent article, \citet{luccioni2024light} argued that users lack critical information and called for more transparency in this regard: \textit{"the majority of users who interact with LLMs have no information about their environmental impacts, and cannot make informed decisions based on model efficiency or carbon intensity"} \cite[][p. 3]{luccioni2025}. This problem has already been addressed in other domains. For example, in the food industry, product labels commonly provide information about ingredients and nutrition \cite{ikonen2020consumer}. In the digital world, labels convey privacy information in apps \cite{li2022understanding}, and content and age ratings in video games \cite{xiao2023shopping} and energy efficiency and carbon labeling are provided for household appliances and vehicles \cite{TormaThogersen2021}. Most relevant for our study are the EU Energy Label \cite{EU2009_125, EU2017_1369} and the US Energy Star Rating. Evidence suggests that labeling approaches help consumers identify better purchase options, including healthier food \cite{polden2025evaluating, ikonen2020consumer} and more efficient household appliances \cite{d2022randomized}. However, with regard to AI, energy labels are still in their infancy. Notable advances include energy labels for open-weight models from Hugging Face\footnote{see \url{https://huggingface.co/AIEnergyScore}}. International organizations such as the OECD \cite{oecd2022measuring} and ongoing regulatory initiatives such as the EU AI Act \cite{EUAIAct2024} have also drawn attention to this issue. However, there is no standard label commonly used in the industry, nor is there any empirical evidence of the effectiveness of such labels. 

\subsection{The Challenge of Energy Disclosure: Performance Trade-offs in AI}
\begin{figure}[h]
  \centering
  \includegraphics[width=1\linewidth]{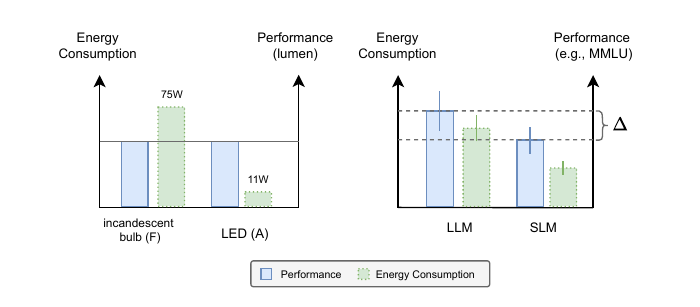}
  \caption{This figure presents an analogy between the luminous efficiency of lightbulbs and the trade-offs in language models. On the left, different lightbulbs produce an identical output (e.g., 1000 lumens) but vary significantly in their energy efficiency. On the right, language models (LLMs and SLMs) differ not only in energy efficiency but also in absolute performance (e.g., MMLU scores of 43.9 vs. 25.9). Critically, unlike the standardized lumen, most users lack the expertise to accurately estimate the true performance difference ($\Delta$) between these models. Error bars indicate variance in energy consumption due to differences in hardware and infrastructure used for inference \cite{jegham2025hungry}.}
  \Description{This figure presents an analogy between the luminous efficiency of lightbulbs and the trade-offs in language models. On the left, different lightbulbs produce an identical output (e.g., 1000 lumens) but vary significantly in their energy efficiency. On the right, language models (LLMs and SLMs) differ not only in energy efficiency but also in absolute performance (e.g., MMLU scores of 43.9 vs. 25.9). Critically, unlike the standardized lumen, most users lack the expertise to accurately estimate the true performance difference ($\Delta$) between these models. Error bars indicate variance in energy consumption due to differences in hardware and infrastructure used for inference \cite{jegham2025hungry}.}
  \label{fig:Comparison}
\end{figure}

Providing users with choices about LLM energy consumption is more complex than it is with other technologies. The energy consumption of lightbulbs offers a useful comparison, given their familiarity and clear evolution from incandescent bulbs to LEDs (Figure \ref{fig:Comparison}, left side). If a user needs to achieve 1000 lumens, the available alternatives differ significantly in their luminous efficiency. An incandescent bulb has very low luminous efficacy and requires $\approx$75 Watts, whereas an LED bulb requires only $\approx$11 Watts for the same output. Therefore, it is easily justifiable to prefer the LED bulb, as no performance trade-off is required.

With LLMs, a user is often confronted with alternatives that also differ significantly in output performance (Figure~\ref{fig:Comparison}, right side). However, since most individuals are not deeply familiar with such technical benchmarks, the true performance difference ($\Delta$) is effectively unknown to them (Figure~\ref{fig:Comparison}, right side). Yet, the labeling of models, such as Gemini PRO versus Gemini Flash or GPT4o versus GPT4o-mini, ensures that most individuals have a vague understanding that models differ in their performance. 

In summary, two important aspects distinguish language models from technologies like lightbulbs from a user's perspective: 
\begin{enumerate}
    \item The performance difference between alternatives is often unknown or difficult to interpret
    \item There is typically a lack of information regarding energy consumption
\end{enumerate}

As a consequence, giving users a choice between AI models presents a considerably more complex decision-making scenario than for established technologies. Therefore, while ECD is a promising concept for AI \cite{luccioni2024light}, its actual impact on user behavior is an open question. Research in other domains suggests that the behavioral effects of energy labels can be surprisingly modest \cite{Waechter2015}. This might also be relevant for language models because of the inherent performance-energy trade-off. This trade-off might not only undermine the goal of the disclosure but could also negatively affect user perception if the sustainable option is seen as simply inferior.

Given the lack of empirical studies exploring this specific dynamic, there is a pressing need to investigate the effectiveness of ECD from a HCI perspective. This need is also echoed in sustainability research, which suggests that more work is required to identify interventions that are truly effective in practice \cite{vanValkengoed2022}. The importance of this research is further intensified by the search for new measures to curb CO\textsubscript{2} emissions. To address this issue, this study is guided by the following two research questions (RQ): 
\newline
\newline
\noindent
$RQ_1$: \textit{To what extent does energy consumption disclosure nudge individuals towards a pro-environmental choice of a language model?}, and \newline 
$RQ_2$: \textit{How does the choice of a model change users' perception of and behavior with the language model?}

\section{Theoretical Foundation}
\subsection{Conceptualization and Research Model}

\begin{figure}[h]
  \centering
  \includegraphics[width=\linewidth]{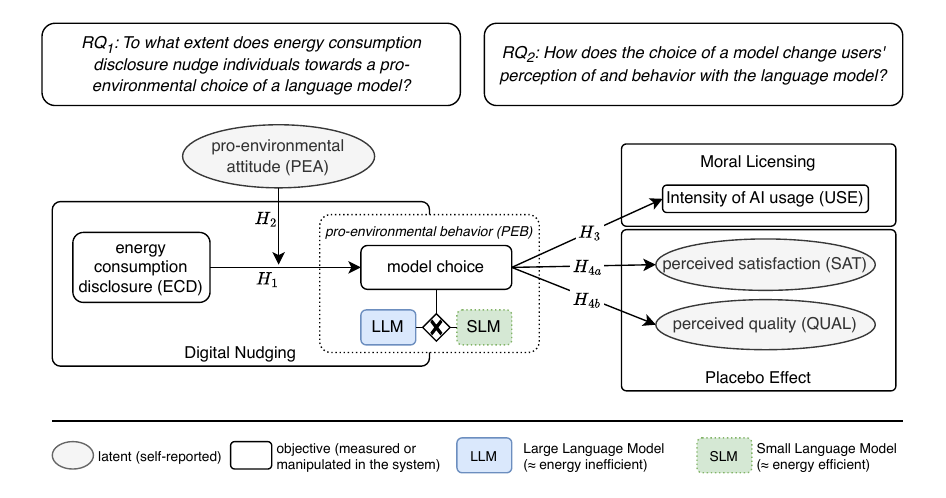}
  \caption{Research Model}
  \Description{Research Model}
  \label{fig:ResearchModel}
\end{figure}

We propose a research model that integrates ECD and PEA to investigate their effects on PEB and three subsequent user outcomes. This model, designed to address $RQ_1$ and $RQ_2$, is depicted in Figure~\ref{fig:ResearchModel}. Table~\ref{tab:constructs} provides an overview of the core constructs and their definitions from the existing literature.

\begin{table*}
  \caption{Core Constructs Used in this Study}
  \label{tab:constructs}
  \begin{tabular}{p{5cm} p{9cm}}
    \toprule
    \textbf{Concept} & \textbf{Definition}\\
    \midrule
    Large Language Model (LLM) & \textit{"advanced language models with massive parameter sizes and exceptional learning capabilities"} \cite[][p. 4]{Chang2024} \\
    Small Language Model (SLM) & language models with fewer parameters and that are designed primarily for efficiency \cite{vannguyen2024surveysmalllanguagemodels}\\
    \midrule
    Energy Consumption Disclosure (ECD)  & information about the energy consumption of the underlying technology \cite{luccioni2024light} \\
    Pro-Environmental Attitude (PEA)  & "\textit{a strong willingness to adopt PEB [pro-environmental behavior] to protect the environment and achieve sustainable development}" \cite[][p. 4]{Shen2024}\\
    Pro-Environmental Behavior (PEB) & \textit{"behaviour that harms the environment as little as possible, or even benefits the environment."} \cite[][p. 309]{Steg2009} \\
    Intensity of AI Usage (USE)  & the extent to which the AI is used \cite{BurtonJones2006}.\\
    Perceived Satisfaction (SAT) & \textit{"the affective attitude towards a specific computer application by someone who interacts with the application directly."} \cite[][p. 261]{Doll1988}\\
    Perceived Quality (QUAL) & \textit{"relates to the extent to which the information provided by the smart product is accurate and reliable."} \cite[][p. 1022]{Chen2024}\\
  \bottomrule
\end{tabular}
\end{table*}

An example of PEB is choosing a "green product" (i.e., products with an eco-friendly image or features) \cite{Schwartz2020} rather than choosing a conventional product. Following the paradigm of the Pro-Environmental Behavior Task \cite{Lange2018}, which uses observed actions rather than self-reported measures, we conceptualize PEB as the user's actual model choice. Therefore, individuals choosing the smaller, less energy-intensive SLM are considered to exhibit PEB. In contrast, those who choose the more energy-intensive LLM do not.

We draw upon \textit{nudge theory} \cite{ThalerSunstein2009} to hypothesize that ECD is an effective means of nudging individuals towards PEB. We use a narrow definition of ECD, recognizing that the energy demand of the underlying LLM is communicated to the user in an aggregated manner. It is noteworthy that industry initiatives, such as the AI Energy Score project from Hugging Face, seek to provide not only detailed information -- such as average inference costs -- but also specifications of the underlying hardware (e.g., NVIDIA H100-80GB). To hypothesize the behavioral and perceptual consequences of model choice, we use \textit{moral licensing theory} \cite{Merritt2010} and research on \textit{placebo effects} in HCI \cite{kosch2023placebo}. To this end, we measure \textit{intensity of AI usage}, \textit{perceived satisfaction}, and \textit{perceived quality} to examine the impact of model choice on these constructs.

\subsection{Effect of Energy Consumption Disclosure on Pro-Environmental Behavior (Model Choice)}

Nudge theory suggests that small changes in the environment or context can influence individuals' behavior \cite{ThalerSunstein2009}. For instance, providing households with information that compares their energy consumption to that of their neighbors typically results in a 2\% decrease in energy use \cite{allcott2011social}. Nudge theory follows the idea of building a "choice architecture," which guides decision-making without restricting individuals' freedom \cite{ThalerSunstein2009}. Nudge theory has been widely applied in fields relevant to our study \cite{Mertens2022}. It has also been argued that nudges are well suited to the digital world ("digital nudging") \cite{Weinmann2016}. Multiple reviews show that nudging is promising for interventions related to the environment. A meta-analysis by \citet{Mertens2022} found that nudges in the environmental domain yield a medium effect size (Cohen's d = 0.4 [0.22; 0.58]). For example, \citet{vonZahn2025} show that green nudging is an effective measure to curtail product returns. This suggests that nudging can be an effective strategy to positively influence individuals' PEB. This is further supported by self-perception theory, which posits that people infer their values from their actions and are motivated to view themselves in a positive light \cite{Schwartz2020}. By making the energy impact of a choice salient, a nudge allows individuals to make a pro-environmental decision, thereby reinforcing a positive, "green" self-perception. Thus, providing energy-related information is expected to encourage pro-environmental actions. However, the effectiveness of nudging is not universal. Recent research highlights that nudges can be ineffective and their outcomes difficult to predict, especially in contexts where other factors are more salient \cite{Szaszi2025}. For example, in a high-stakes situation, a user's primary goal might override the subtle influence of an energy disclosure nudge. As a result, \citet{Szaszi2025} advise expecting only small average effects from these types of interventions. Recognizing these limitations, the established effectiveness of informational nudges in the environmental domain provides a strong basis for investigation in the novel context of LLM interactions. While effects may be modest, the expected direction is positive. Therefore, we propose the following hypothesis (H):
\newline
\noindent
\textit{$H_1$: Disclosing the energy consumption of a language model positively influences a user's choice towards more energy-efficient options.}
\newline
\newline
The effectiveness of nudge interventions is often moderated by individual differences \cite{Ghesla2020, Kellen2021}. Within the sustainability domain, PEA, in particular, has been identified as a fundamental predictor of behavior \cite{Shen2024}. Here, we also assume that individual attitude is an important precursor to PEB. Consequently, it is reasonable to expect that the effect of ECD on PEB ($H_1$) will also be moderated. Given its established importance, we posit that an individual's pre-existing PEA will influence the extent to which the nudge is effective. More specifically, we expect the positive effect of the energy disclosure nudge to be stronger for individuals with a higher PEA. Conversely, the nudge is expected to have a weaker effect on individuals with low PEA, as the information may not align with their existing values. Therefore, we propose our second hypothesis:
\newline
\noindent
\textit{$H_2$: Pro-environmental attitude moderates the relationship between energy consumption disclosure and pro-environmental behavior, such that the positive effect of the disclosure is stronger for individuals with higher levels of pro-environmental attitude.}

\subsection{Consequences of Model Choice on Use Behavior}
Extant literature implies that interventions aiming at PEB can cause spillover effects. A spillover occurs when an intervention targeting one behavior influences subsequent behaviors that were not the primary focus \cite{truelove2014positive}. These effects can be positive, where a specific type of PEB encourages another type of PEB. For example, individuals who purchase an electric vehicle are also more likely to engage in other sustainable energy behaviors \cite{Peters2018}. Alternatively, spillover effects can be negative, such as when one PEB discourages others. For instance, a study by \citet{barr2010holiday} found that environmentally conscious behavior at home morally freed people from the need to be environmentally friendly while on vacation. This caused environmentally conscious individuals to be the most likely to use the most carbon-intensive mode of transportation. These effects are also known as rebound and moral licensing effects. The rebound effect, which is often analyzed by economists in the context of Jevons' paradox \cite{alcott2005jevons}, is typically linked to gains in efficiency. It occurs when increased efficiency lowers the cost of a resource, leading to higher overall consumption. For instance, when a vehicle becomes more fuel-efficient and less expensive to operate, its owner may drive it more frequently. This concern also applies to AI, where improved hardware efficiency for training and operating models could paradoxically lead to increased demand and energy use \cite{luccioni2025efficiency}. Similarly, moral licensing, which stems from social psychology \cite{blanken2015meta}, posits that individuals who perform a morally positive act may subconsciously feel justified in making a less virtuous choice later. The core idea is that people manage a "moral self-concept." Performing a PEB enhances this concept, creating a surplus of moral value. This surplus can make them feel "off the hook" and allow them to engage in subsequent, less ethical behaviors \cite{blanken2015meta, mullen2016consistency, khan2006licensing}. For instance, \citet{klockner2013positive} showed that owners of electric cars can feel they have already offset their contribution to pollution. This belief can reduce their guilt associated with using the car, potentially leading to more frequent use.

Our study investigates how disclosing an AI model's energy consumption information influences user behavior. Drawing from the concept of moral licensing, we hypothesize that users who choose a more energy-efficient AI model may feel a reduced moral obligation to engage in subsequent pro-environmental actions. In the context of AI use, these behaviors include the frequency and length of user-model interactions. We therefore expect that choosing an SLM, the energy-efficient model, will lead users to engage in a greater number of interactions or use longer prompts. Thus:
\newline
\noindent
\textit{$H_3$: Choosing a lower performance model leads to a more intensive use of AI.}

\subsection{Consequences of Model Choice on Perception}
Users often encounter different models that vary in performance for a given task. This raises the question of whether users, particularly non-experts, can reliably discern differences in model performance. In settings where quality is subjective or difficult to validate, the concept of the \textit{placebo effect} has been observed in various scenarios. The placebo effect is a psychological phenomenon in which an inert procedure or substance, which has no inherent power, elicits a real, observable outcome \cite{stewart2004placebo}. The classic example of this phenomenon comes from medicine. A patient's health can improve after receiving a placebo (a pill with no effect) simply due to the belief that it is a potent pharmaceutical \cite{stewart2004placebo}. This effect is largely driven by expectation: the recipient's belief that a particular outcome will occur can help bring about that outcome \cite{price2008comprehensive}. The placebo effect is commonly observed in medicine \cite{levine1978mechanism} and psychology \cite{price2008comprehensive}, but it has also recently been observed in HCI, especially in AI systems \cite{kosch2023placebo, vaccaro2018illusion, denisova2015placebo}. For example, \citet{denisova2015placebo} demonstrated that priming players to expect an adaptive AI in a game positively influenced their experience, even when the AI was not actually adaptive. Similarly, \citet{vaccaro2018illusion} found that users reported greater satisfaction with social media news feeds when offered a control setting, regardless of whether the setting was functional. More recently, \citet{kosch2023placebo} informed one group of participants that a system supporting a word puzzle task was adaptive and told another group that it was not. In reality, neither system was adaptive. However, those who were led to believe they were working with an adaptive system reported higher performance expectancy.

Major LLM providers, such as OpenAI and Google, often indicate performance differences through designated model tiers, such as "pro," "mini," or "flash." Users may also encounter more explicit performance indicators beyond these marketing labels. These cues likely shape user expectations before an interaction begins. In this study, we argue that the placebo effect is a key concept to consider because users' subjective evaluations are susceptible to expectation setting. Specifically, we hypothesize that a model's advertised performance influences user perception and satisfaction. We predict that users who select a model with lower performance (which, in our study, is linked to greater energy efficiency) will report lower interaction quality and satisfaction than users who select a high-performing model, even if the underlying technology is the same. Therefore, we propose our final hypotheses:
\newline
\noindent
\textit{$H_4a$: Choosing a lower performance model leads to a lower degree of satisfaction.}
\newline
\noindent
\textit{$H_4b$: Choosing a lower performance model leads to a lower degree of perceived quality.}

\section{Methodology}
\subsection{Experimental Design}
We conducted a one-factorial between-participants online experiment to investigate the impact of exposure to ECD on user behavior (model choice). The single factor was the presence of ECD, with two levels: a treatment condition in which participants received information about the energy use and performance of two different language models, and a control condition in which only the performance information was shown (see Section \ref{sec:manipulation}). To isolate the psychological effects of choice and expectation, both the high-performance "LLM" and the energy-efficient "SLM" were, unbeknownst to participants, powered by the same underlying model (GPT-4o mini). This design allowed us to test for a placebo effect, in which perceived differences were based on framing rather than objective performance. All data and the corresponding analysis files are accessible on the Open Science Framework\footnote{\url{https://osf.io/mvjhy}}.

\subsection{Participants}
We conducted a power analysis to determine the required sample size to detect small to medium-sized effects $w=0.15$ \cite{Mertens2022} with a power of 80\% and $\alpha = 0.05$, which suggested a minimum sample size of 349. Participants were recruited from the crowdsourcing provider Prolific \cite{albert2023comparing} as crowd workers widely use LLMs \cite{veselovsky2023artificialartificialartificialintelligence}. We recruited a total of 367 participants who successfully completed the experiment. To enhance data quality, we excluded observations from participants who indicated that their answers were not honest, yielding \emph{N}\,=\,365 observations used for the analysis. Each participant received £1.00 ($\approx$ \$1.33) for a mean participation time of 9 minutes and 56 seconds. We used custom screening to ensure that the sample aligned with our study purpose. Specifically, we only included individuals who were born in the United Kingdom (UK), had English as their first language, had UK nationality, and had a previous approval rate between 99\% and 100\%. Both full-time and part-time employees were eligible to participate. We chose UK citizens because they are familiar with the European Energy Label, which we use as reference for our study. An overview of the demographic characteristics is shown in Table \ref{tab:demographics}. Section \ref{sec:appendix_randomization} in the Appendix shows that randomization was successful.

\begin{table*}
\caption{Demographic Characteristics}
\label{tab:demographics}
\begin{tabular}{lcccccc}
\toprule
 \textbf{Category} &  \textbf{n} & \textbf{\%} & \textbf{Control} & \textbf{Control (\%)}  & \textbf{Treatment} & \textbf{Treatment (\%)} \\
\midrule
 \multicolumn{7}{l}{Gender} \\
 \hspace{5mm}Woman & 166 & 45.6 & 85 & 44.5 & 81 & 46.8 \\
 \hspace{5mm}Man & 194 & 53.3 & 104 & 54.5 & 90 & 52.0 \\
 \hspace{5mm}Genderqueer & 1 & 0.3 & 1 & 0.5 & 0 & 0.0 \\
 \hspace{5mm}Nonbinary & 1 & 0.3 & 0 & 0.0 & 1 & 0.6 \\
 \hspace{5mm}Prefer not to say & 2 & 0.5 & 1 & 0.5 & 1 & 0.6 \\
\midrule
 \multicolumn{7}{l}{Age} \\
 \hspace{5mm}18--24 & 22 & 6.0 & 7 & 3.6 & 15 & 8.7 \\
 \hspace{5mm}25--34 & 96 & 26.3 & 51 & 26.6 & 45 & 26.0 \\
 \hspace{5mm}35--44 & 94 & 25.8 & 50 & 26.0 & 44 & 25.4 \\
 \hspace{5mm}45--54 & 91 & 24.9 & 47 & 24.5 & 44 & 25.4 \\
 \hspace{5mm}55--64 & 53 & 14.5 & 30 & 15.6 & 23 & 13.3 \\
 \hspace{5mm}65--74 & 9 & 2.5 & 7 & 3.6 & 2 & 1.2 \\
 \midrule
 \multicolumn{7}{l}{Education} \\
 \hspace{5mm}High school diploma or equivalent & 60 & 16.6 & 39 & 20.6 & 21 & 12.2 \\
 \hspace{5mm}Some college, no degree & 48 & 13.3 & 17 & 9.0 & 31 & 18.0 \\
 \hspace{5mm}Associate degree & 19 & 5.3 & 8 & 4.2 & 11 & 6.4 \\
 \hspace{5mm}Bachelor's degree & 154 & 42.7 & 84 & 44.4 & 70 & 40.7 \\
 \hspace{5mm}Master's degree & 56 & 15.5 & 31 & 16.4 & 25 & 14.5 \\
 \hspace{5mm}Doctoral or professional degree & 20 & 5.5 & 7 & 3.7 & 13 & 7.6 \\
 \hspace{5mm}Prefer not to say & 4 & 1.1 & 3 & 1.6 & 1 & 0.6 \\
\bottomrule
\end{tabular}
\end{table*}

\subsection{Implementation of the Experiment}
\label{sec:manipulation}
We implemented an online experimental environment using Node.js and Next.js for the backend, Bootstrap CSS for the front-end, and Survey.js to include a questionnaire after the experiment. The experimental manipulation involves a choice architecture \cite{ThalerSunstein2009}. Participants in the treatment condition were asked to choose between two fictional models, both named "GREGALE," to mitigate brand-related biases associated with commercial models. To avoid unwanted priming effects, we did not use commercial qualifiers such as "flash vs. pro" or reveal model parameters to participants. We refer to these two models as (1) LLM and (2) SLM based on their performance characteristics, although these terms were not shown to participants. Recall that both the LLM and SLM models were powered by the same underlying GPT-4o mini system. Inspired by information provided in current leaderboards\footnote{\url{https://lmarena.ai/leaderboard/text}}, we presented aggregated performance information for both models, including the scoring date (April 2025), domain (text generation), and a star-rating performance metric.

For the LLM model, we used a five out of five performance rating ({\raisebox{-1.9pt}{\fontsize{13pt}{12pt}\ding{72}}}{\raisebox{-1.9pt}{\fontsize{13pt}{14pt}\ding{72}}}{\raisebox{-1.9pt}{\fontsize{13pt}{14pt}\ding{72}}}{\raisebox{-1.9pt}{\fontsize{13pt}{14pt}\ding{72}}}{\raisebox{-1.9pt}{\fontsize{13pt}{14pt}\ding{72}}}). In contrast, the SLM had a three out of five rating ({\raisebox{-1.9pt}{\fontsize{13pt}{14pt}\ding{72}}}{\raisebox{-1.9pt}{\fontsize{13pt}{14pt}\ding{72}}}{\raisebox{-1.9pt}{\fontsize{13pt}{14pt}\ding{72}}}{\raisebox{-1.9pt}{\fontsize{13pt}{14pt}\ding{73}}}{\raisebox{-1.9pt}{\fontsize{13pt}{12pt}\ding{73}}}). This performance difference was designed to reflect the relative performance of current LLMs (e.g., GPT-4o vs. GPT-4o-mini). In the treatment condition, we introduced additional information: an energy efficiency score. This score was presented using a label similar to the European Union's energy label, ranging from \textbf{A} (very positive) to \textbf{G} (very poor). The LLM received an \textbf{F} score, whereas the SLM received a \textbf{B} score. The control condition included only the performance star ratings. All other information was kept identical across both conditions. An overview of the information presented is shown in Figure \ref{fig:manipulation}. Details on the experimental procedure and the measurement instrument are provided in Appendix \ref{sec:appendix_experimental_procedure} and \ref{sec:appendix_measurement_instrument}.


\begin{figure}[h]
  \centering
  \includegraphics[width=\linewidth]{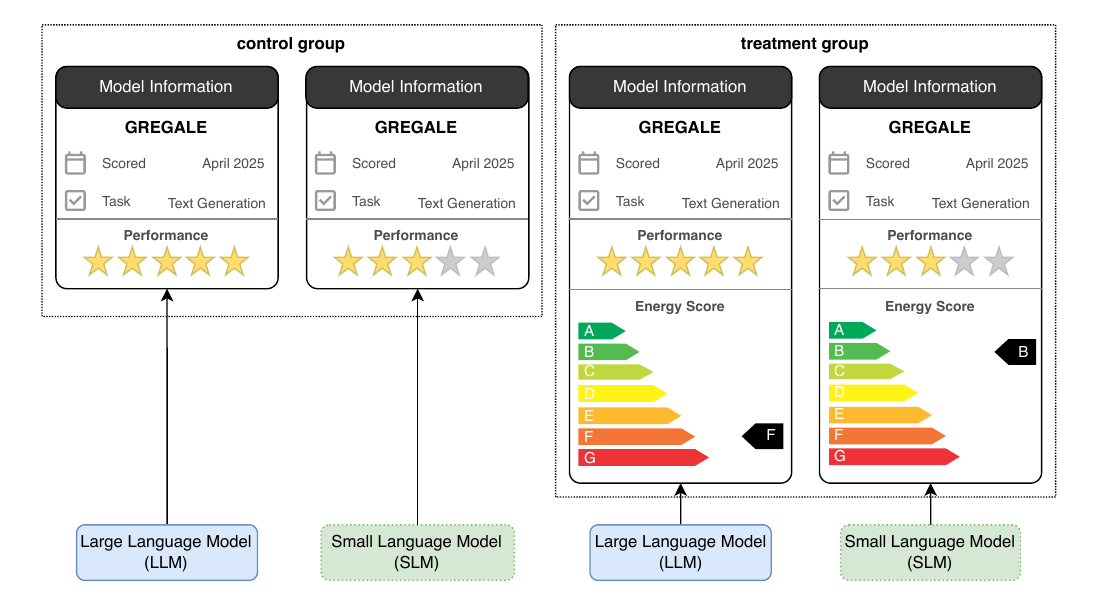}
  \caption{The experimental manipulation shown to participants. The control condition (left) displayed only performance ratings, while the treatment condition (right) included an energy efficiency score (A--G), creating a choice between performance and sustainability.}
  \Description{The experimental manipulation shown to participants. The control condition (left)  displayed only performance ratings, while the treatment condition (right) included an energy efficiency score (A--G), creating a choice between performance and sustainability.}
  \label{fig:manipulation}
\end{figure}

\section{Results}
\begin{figure}[h]
  \centering
  \includegraphics[width=.8\linewidth]{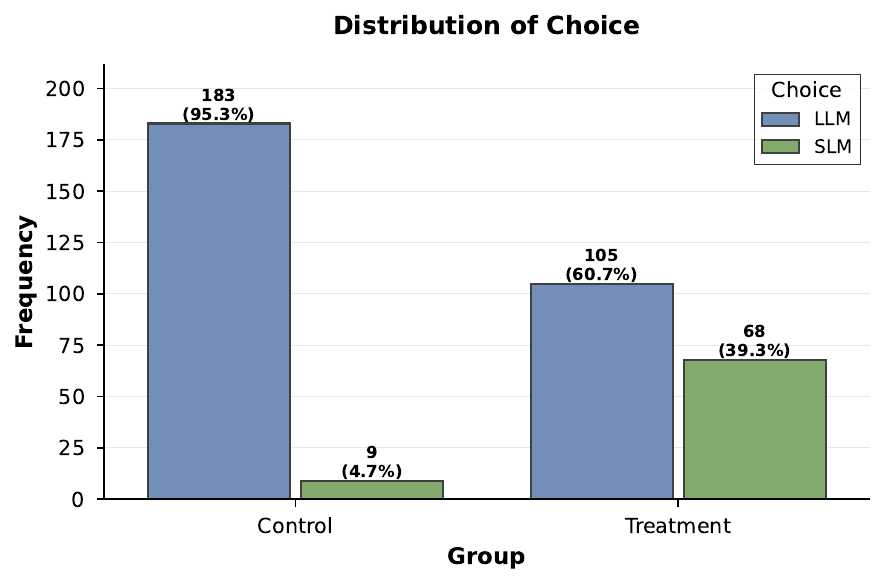}
  \caption{ECD altered the choices made by participants. Count plot illustrating the number of individuals in the control (n = 192) and treatment (n = 173) groups who selected either the LLM or the SLM. The distribution of choices between the two groups was statistically significant ($\chi^2$(1, 365) = 269.63, p < .0001).}
  \Description{ECD altered the choices made by participants. Count plot illustrating the number of individuals in the control (n = 192) and treatment (n = 173) groups who selected either the LLM or the SLM. The distribution of choices between the two groups was statistically significant ($\chi^2$(1, 365) = 269.63, p < .0001).}
  \label{fig:ChoiceDistribution}
\end{figure}

To test $H_1$, we conducted a $\chi^2$ test to investigate the effect of the manipulation on model choice. Our results show a significant effect of the manipulation and an increase in selection of the SLM model in the treatment group ($\chi^2$(1, 365) = 269.63, p < .0001). In the control group, only 4.7\% of participants selected the SLM model. In contrast, 39.3\% preferred the SLM model over the LLM model in the treatment condition. An overview of the distribution of model choice is shown in Figure \ref{fig:ChoiceDistribution}.

To test $H_2$, we conducted a binary logistic regression to examine the effect of condition (ECD), PEA, and their interaction on the likelihood of model choice. The model included 365 participants and was statistically significant, $\chi^2(3) = 89.28, p < .001$, indicating that the predictors distinguished between outcome categories. The model explained approximately 23.7\% of the variance in choice (McFadden’s pseudo $R^2 = 0.2374$). Treatment had a positive, strong, and significant effect on model choice ($B = 2.56$, $SE = 0.38$, $z = 6.72$, $p < .001$), meaning that participants in the treatment condition were more likely to choose the SLM model than those in the control condition. The main effect of PEA score was not significant ($B = -0.16$, $SE = 0.33$, $z = -0.48$, $p = .631$). However, the interaction between condition and PEA score was significant ($B = 0.90$, $SE = 0.38$, $z = 2.36$, $p = .018$), indicating that the effect of condition on model choice varied as a function of PEA. Specifically, the effect of condition on choice was stronger for participants with higher levels of PEA. These results suggest that ECD influences model choice, and the strength of the effect depends on the user's level of PEA. Regarding the odds ratio (OR), we observe that exposure was associated with increased odds of the outcome (OR = 12.89; 95\% CI: 6.11 -- 27.17). Table \ref{tab:logit} provides an overview of the logistic regression results. 

\begin{table*}[]
\centering
\caption{Binary Logistic Regression Predicting Model Choice}
\label{tab:logit}
\begin{tabular}{p{3.5cm}ccccccc}
\toprule
 & \textbf{Coefficient} & \textbf{Std. Error} & \textbf{z-value} & \textbf{p-value} & \textbf{2.5\% CI} & \textbf{97.5\% CI} & \textbf{Odds Ratio} \\
\midrule
Intercept & -3.01 & 0.34 & -8.79 & 0.00 & -3.68 & -2.34 & 0.05 \\
ECD & 2.56 & 0.38 & 6.72 & 0.00 & 1.81 & 3.30 & 12.89 \\
PEA & -0.16 & 0.32 & -0.48 & 0.63 & -0.79 & 0.48 & 0.86 \\
ECD $\times$ PEA & 0.90 & 0.38 & 2.36 & 0.02 & 0.15 & 1.65 & 2.46 \\
\bottomrule
\end{tabular}
\parbox{\linewidth}{\footnotesize \textit{Note.} Coefficients are unstandardized logistic regression estimates. The dependent variable is model choice (0 = LLM, 1 = SLM).}
\end{table*}

\begin{figure}[h]
  \centering
  \includegraphics[width=\linewidth]{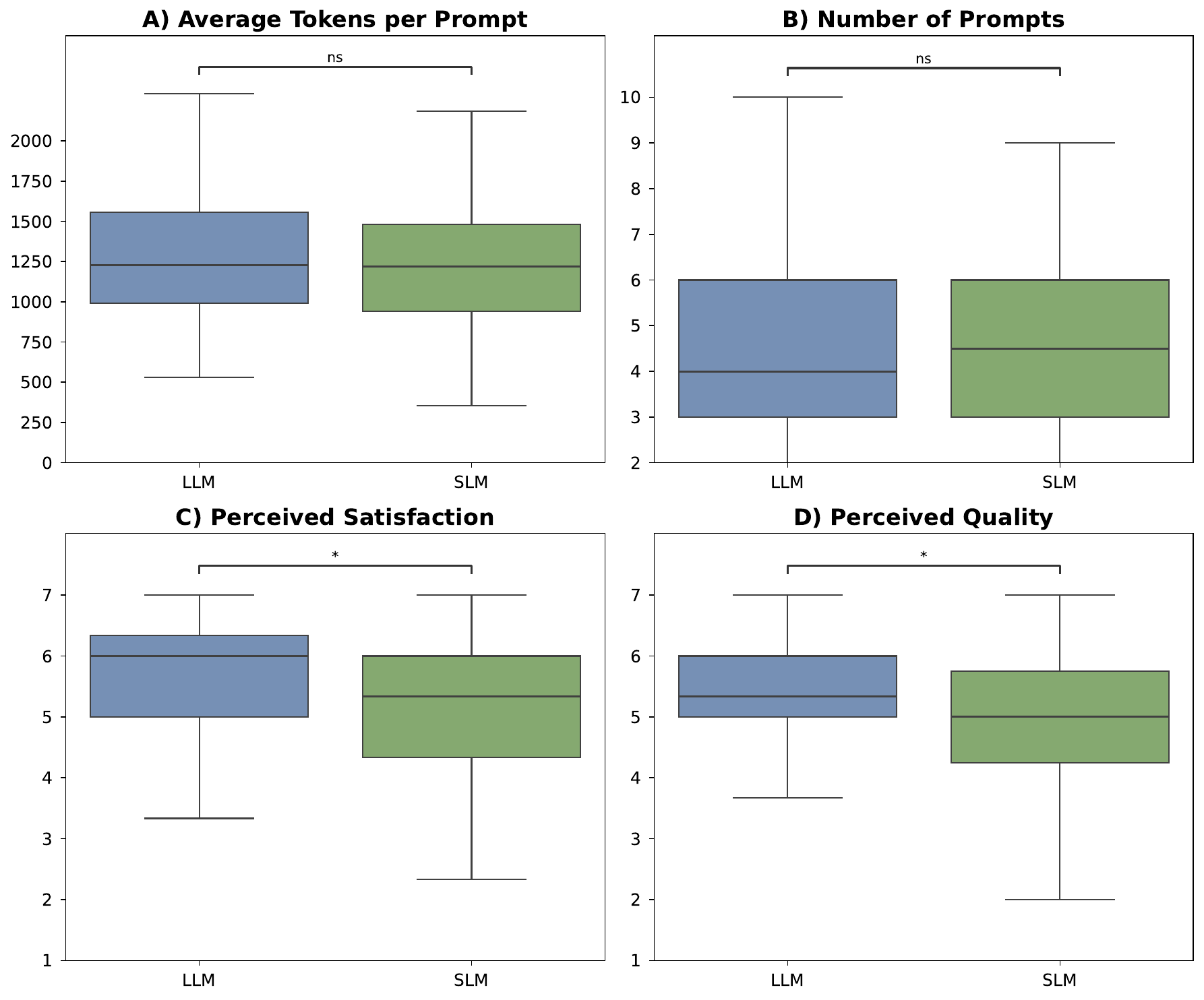}
  \caption{Differences in behavior and perception based on model choice. The results are based on \emph{N} = 173 observations that received the treatment. The figure shows four boxplots comparing key dependent variables for participants grouped by their choice of the LLM (n = 105) versus the SLM (n = 68). The variables displayed are (A) Average Tokens per Prompt, (B) Number of Prompts, (C) Perceived Satisfaction, and (D) Perceived Quality. Pairwise comparisons were conducted using Mann--Whitney U tests with Holm--Bonferroni correction. Significance levels are denoted as: $^*p \leq 0.05$, $^{**}p \leq 0.01$, $^{***}p \leq 0.001$, $^{****}p \leq 0.0001$.}
  \Description{Differences in behavior and perception based on model choice. The results are based on \emph{N} = 173 observations that received the treatment. The figure shows four boxplots comparing key dependent variables for participants grouped by their choice of the LLM (n = 105) versus the SLM (n = 68). The variables displayed are (A) Average Tokens per Prompt, (B) Number of Prompts, (C) Perceived Satisfaction, and (D) Perceived Quality. Pairwise comparisons were conducted using Mann--Whitney U tests with Holm--Bonferroni correction. Significance levels are denoted as: $^*p \leq 0.05$, $^{**}p \leq 0.01$, $^{***}p \leq 0.001$, $^{****}p \leq 0.0001$.}
  \label{fig:boxplots_ttest}
\end{figure}

Next, to test $H_3, H_{4a}$, and $H_{4b}$, we use a sub-sample consisting only of participants from the treatment group ($N=173$) to reduce issues related to self-selection bias \cite{Heckman1979}. We conducted a series of Mann--Whitney U tests to examine the effect of the independent variable, model choice, on the four dependent variables: \emph{average tokens per prompt}, \emph{number of prompts}, \emph{perceived satisfaction}, and \emph{perceived quality}. A Holm--Bonferroni correction was applied to control for the family-wise error rate across the multiple tests \cite{Holm1979}. Our results show no significant difference in average tokens per prompt between the LLM group (Mdn = 1228.50, n = 105) and the SLM group (Mdn = 1220.46, n = 68), \emph{U} = 3787.0, \emph{p-adjusted} = 1.000, \emph{r} = 0.06. Similarly, we observe no significant difference in number of prompts between the LLM group (Mdn = 4.00, n = 105) and the SLM group (Mdn = 4.50, n = 68), \emph{U} = 3551.5, \emph{p-adjusted} = 1.000, \emph{r} = -0.01. In contrast, our results show a significant difference in \emph{perceived satisfaction} between the LLM group (Mdn = 6.00, n = 105) and the SLM group (Mdn = 5.33, n = 68), \emph{U} = 4487.0, \emph{p-adjusted} = 0.017, \emph{r} = 0.26, and a significant difference in \emph{perceived quality} between the LLM group (Mdn = 5.33, n = 105) and the SLM group (Mdn = 5.00, n = 68), \emph{U} = 4520.5, \emph{p-adjusted} = 0.012, \emph{r} = 0.27. An overview of the effect of model choice on our dependent variables is shown in Figure \ref{fig:boxplots_ttest}. A post-hoc power analysis suggests very low power for \emph{average tokens per prompt} (Power = 0.122) and \emph{number of prompts} (Power = 0.051) and high power for \emph{perceived satisfaction} (Power = 0.924) and \emph{perceived quality} (Power = 0.942).    

\section{Discussion}
\subsection{Summary of Key Findings}
The rapid growth in the size and complexity of language models has led to a significant increase in their energy consumption and resource demands. Addressing this environmental footprint is crucial for the sustainable development of AI. International organizations (e.g., OECD) and regulators (e.g., the EU) have acknowledged this issue and are encouraging greater transparency regarding AI energy consumption \cite{oecd2022measuring, EUAIAct2024}. A promising avenue is to enhance transparency on the client side, empowering users to make environmentally informed decisions \cite{luccioni2024light, luccioni2025, Luccioni2023BLOOM}. This study is guided by two research questions and four corresponding hypotheses (Table~\ref{tab:summaryhypothesis}).

With regard to our first research question (\textit{RQ\textsubscript{1}: To what extent does energy consumption disclosure nudge individuals towards a pro-environmental choice of a large language model?}), we found this intervention to be highly effective. The presence of an energy label increased the selection of a smaller, more efficient model (SLM) from just 4.7\% in the control group to 39.3\% in the treatment group. This corresponds to an OR greater than 12, indicating that participants odds of choosing the SLM (vs. the LLM) were more than 12 times higher when energy efficiency was disclosed. Furthermore, this effect was significantly moderated by PEA, proving to be more pronounced among individuals with a greater awareness of the environment.

With regard to our second research question (\textit{RQ\textsubscript{1}: How does the choice of a model change users' perception of and behavior with the language model?}), we found mixed results. Specifically, while we found no evidence of a moral licensing effect -- whereby users who make a pro-environmental choice feel licensed to behave less pro-environmentally later -- we uncovered a critical unintended consequence. Participants who chose the energy-efficient SLM, which was framed as less powerful, subsequently reported lower satisfaction and perceived quality of interaction. We identify this as a placebo effect: the mere expectation of lower performance negatively influenced their subjective experience with the system. This finding highlights a significant challenge in promoting sustainable AI, as the framing required to encourage adoption may inadvertently degrade user perception.

In summary, our study shows that fostering PEB in the context of language models comes with a trade-off. On the positive side, it is possible to enhance PEB significantly using nudging; on the other hand, there is also a risk of fostering unintended consequences, such as negative perceptions. 

\begin{table*}[]
\centering
\caption{Overview Hypothesis}
\label{tab:summaryhypothesis}
\begin{tabular}{@{}p{0.5cm}p{12.5cm}c@{}}
  \toprule
  \textbf{Hypothesis} & & \textbf{Result} \\
  \midrule
  $H_1$ & Disclosing the energy consumption of a language model positively influences a user's choice towards more energy-efficient options. & \cmark \\
  $H_2$ & Pro-environmental attitude moderates the relationship between energy consumption disclosure and pro-environmental behavior, such that the positive effect of the disclosure is stronger for individuals with higher levels of pro-environmental attitude. & \cmark \\
  $H_3$ & Choosing a lower performance model leads to a more intensive use of AI. & \xmark \\
  $H_{4a}$ & Choosing a lower performance model leads to a lower degree of satisfaction. & \cmark \\
  $H_{4b}$ & Choosing a lower performance model leads to a lower degree of perceived quality. & \cmark \\
  \bottomrule
\end{tabular}
\end{table*}

\subsection{Implications for Theory}
Nudge theory provides a robust theoretical lens for designing behavioral interventions \cite{ThalerSunstein2009, Mertens2022}. In our study, we used odds ratios to quantify the effect size, as they are a straightforward metric \cite{gotz2024effect}. We interpret our odds ratio of over 12 as a substantial effect, far exceeding the modest outcomes typically reported in the nudging literature \cite{Szaszi2025}. From a theoretical standpoint, these results demonstrate that it is possible to design nudges with a powerful impact on pro-environmental choices, at least in controlled settings. This strong effect suggests that energy labels could remain effective when deployed in real-world scenarios, though likely with a diminished impact. 

Our findings also contribute to the ongoing discussion about the generalizability of nudges \cite{Szaszi2025}. The literature reminds us that nudges are not universally effective and that we must account for heterogeneity among individuals. By identifying PEA as a key moderator, we show significant variation in the effectiveness of our nudge. Future research aiming to apply nudging for sustainability can draw on this insight by incorporating PEA to better predict and segment user behavior.

PEB is a pivotal construct for research on sustainability. This is one of the first studies to use AI model choice as a specific type of pro-environmental task, extending the work of \cite{Lange2018} to an HCI context. Despite mixed results in previous literature, we hypothesized that PEA would be significantly related to PEB, and our results confirm a significant moderation effect. In the context of our study, we show there is a significant moderation effect, which differs from the results of \cite{Shen2024}, who did not find a significant relationship. This outcome may be linked to differences in data collection methodology. While some studies draw on survey data \cite{Shen2024}, we employed an experimental design that measured actual behavior with objective data, followed by a post-questionnaire. Our findings suggest that attitudes are more closely correlated when responses are collected after a specific task than in research settings where individuals provide answers to a survey.

We use the concept of the moral licensing effect \cite{blanken2015meta} to investigate potential unintended negative consequences. To our surprise, we did not observe a significant difference in subsequent behaviors, such as the number of prompts or the average tokens per prompt (see Figure \ref{fig:boxplots_ttest}). While the differences were not statistically significant, we did observe a slight variation in the number of prompts. We speculate that this minor difference could become significant in other contexts. For example, increasing task complexity, such as requesting detailed summaries or well-formatted documents, would naturally increase the number of prompts needed, potentially amplifying this effect. Future research investigating moral licensing in the context of AI should consider manipulating task complexity to gain deeper insights into these potential consequences.

An increasing body of knowledge in HCI has provided insights into the placebo effect \cite{kosch2023placebo, vaccaro2018illusion, denisova2015placebo}, and our findings align with this research. We show that the framing of a model's performance creates a placebo effect in which the expectation of lower quality leads to a lower perceived quality. This phenomenon presents both an opportunity and a risk. On the one hand, it suggests that the framing of a model may be just as important as its underlying technical performance. On the other hand, being transparent about a model's trade-offs (e.g., lower performance for better energy efficiency) can negatively impact the user experience. This creates a dilemma, as the push for sustainable AI requires transparency. More research is needed to explore how to inform users about performance and energy consumption while minimizing negative placebo effects. For instance, future studies could test whether changing the nudge's format from an aggregated rating (e.g., {\raisebox{-1.9pt}{\fontsize{13pt}{14pt}\ding{72}}}{\raisebox{-1.9pt}{\fontsize{13pt}{14pt}\ding{72}}}{\raisebox{-1.9pt}{\fontsize{13pt}{14pt}\ding{72}}}{\raisebox{-1.9pt}{\fontsize{13pt}{14pt}\ding{73}}}{\raisebox{-1.9pt}{\fontsize{13pt}{12pt}\ding{73}}}) to an absolute performance metric (e.g., MMLU = 65.9) could help manage expectations while remaining fully transparent.


\subsection{Implications for Design}
Our findings offer several implications for designing user interfaces and choice architectures that promote sustainable AI interaction. First and foremost, our study shows how to use ECD effectively to design user interfaces fostering PEB. This contributes to previous HCI literature which has shown that PEB can be positively influenced using gamification elements \cite{vanhoudt2020} and that eco-feedback can positively influence energy-saving behavior \cite{Berney2024}. Our design is informed by nudge theory, which allows users to maintain freedom of choice while still enhancing levels of PEB. Another advantage of this approach lies in its simplicity. Designers only need to add labels similar to the ones used here (see Figure \ref{fig:manipulation}) in existing model selection interfaces. 

Second, our findings highlight a significant "Good for the Planet, Bad for Me" challenge, requiring designers to move beyond simple nudges and develop more sophisticated interventions to navigate this trade-off. For instance, designers could emphasize performance gains, such as faster response times, to position an SLM more favorably and reduce the perceived compromise for users. Similarly, they could add qualitative information about an SLM's strengths, such as "this model tends to be concise" or "this model is optimized for creative work." In doing so, designers can mitigate the perceptual downgrading associated with choosing an SLM. Unlike our study, which used a single model to investigate a placebo effect, real-world organizations implement various models to leverage the distinct benefits of both SLMs and LLMs. Consequently, users perceive actual performance differences, which could result in more pronounced outcomes than those observed in our study, which were based primarily on expectations. This makes such mitigation strategies even more crucial for organizations.

There is ample evidence that there is no "one-size-fits-all" approach to designing user interfaces \cite[e.g.,][]{Ding2024}. Here, we demonstrate that designers face challenges when accounting for individual differences to increase the impact of ECD. For instance, designers could use tailored nudges to engage users who are less motivated by environmental concerns. For users with low PEA levels, designers can provide information about cost reductions through lower API costs. This approach allows the system to appeal to a broader range of users.

Third, our results indicate that designers may want to consider alternative methods for encouraging users to promote PEB. One of the most effective nudges is the default setting -- an opt-out system where the target option is set as the default \cite{Ning2022}. To leverage the strength of the default option, designers can set the SLM as default to further enhance the adoption of the pro-environmental option. In this regard, our study can be considered a point of departure for adjusting the way ECD is conveyed to users.  

Finally, designers might also consider avoiding burdening users with this choice, and instead automatically routing queries to the most efficient model capable of handling a request. This concept -- known as dynamic model routing \cite{Google2025ModelRouting} -- is a powerful approach for selecting the smallest feasible language model based on characteristics such as task complexity or prompt complexity. Designers could also give informed users the option to override the automatic process, allowing them to maintain autonomy.

\subsection{Implications for Society}
Based on our results, we assume that the widespread use of energy labels in AI systems, which is uncommon today, has the potential to influence individual choices and increase awareness of the topic. Informing and educating users is an important prerequisite for societal change. A recent meta-analysis suggests that climate change education \cite{AeschbachSchwichowRiess2025} -- and ECD is a subtle form of climate change education -- is an effective means of influencing knowledge, attitudes, and behavior. User education is critical, as it has been identified as a key factor in broader efforts to decarbonize digital infrastructures, such as data centers \cite{Liu2025}. Using ECD as explored here is, therefore, an important foundation for promoting societal change toward greater sustainability. 

Although our research focused on individual behavior, we argue that the findings hold significant implications for subsequent collective action for two primary reasons. First, ECD is intended for end users rather than developers \cite{Goercu2025} which means that it has the potential to impact a very large number of people. Second, previous literature highlights that environmental awareness and clear information are pivotal drivers of both individual and collective action in support of sustainable use \cite{Walters2025}. Therefore, individuals who are exposed to ECD are likely to become more reflective about the energy consumption of AI. Being more reflective about model choice may ultimately lead to spill-over effects within one's peer group, thus serving as a trigger for collective action.

\subsection{Limitations and Future Work}
While our findings provide compelling evidence for the effectiveness of ECD as a nudging mechanism, our study has several limitations that constrain the generalizability of the results and suggest avenues for future research. 

First, our experimental design includes a single, low-stakes task. Therefore, the perceived trade-off between model performance and sustainability may be significantly different in high-stakes situations. This includes medical, legal, or financial decisions, where accuracy is critical and the consequence of an inferior response are far-reaching. Therefore, we caution that the findings of this study cannot be generalized to these domains and require future research to explore the impact of ECD in these scenarios. 

Second, we used a simple, aggregated design to inform participants about energy consumption and performance, combining a star rating for performance with an energy label similar to EU-labels. This approach does not allow us to disentangle the individual effects of performance cues and energy labels, nor does it explore the broader design space of disclosure. Therefore, future work should systematically investigate alternative visualizations and framings. This could include using more concrete metrics (e.g., carbon emissions in grams or ELO scores), framing the cost in more relatable terms (e.g., "this query is equivalent to 0.5 W"), or comparing static versus dynamic labels that update in real time based on query complexity.

Third, our observation of a placebo-like effect on perceived satisfaction and perceived quality is significant, but our study does not examine how this bias might be mitigated. As we have identified a key dilemma for designers aiming to foster pro-environmental choices at the cost of user perception, a critical next step is to develop and evaluate design interventions that explicitly address this trade-off. This could involve exploring alternative choice architectures or highlighting the co-benefits of sustainable options, such as faster response times or cost savings.

Fourth, we included only individuals from the UK. Therefore, we cannot draw conclusions beyond this sample. Other countries use different types of energy labels, which may lead to varying levels of familiarity and effects compared to those observed here. We encourage future research to examine the effectiveness of energy labels in cross-cultural settings. 

Fifth, our experiment was conducted in a single, short-term session. While this controlled setting was ideal for isolating choice behavior and drawing conclusions about the effectiveness of ECD, it does not allow us to draw conclusions about long-term effects. The absence of a moral licensing effect in our data may simply be a consequence of the design. Future research should employ longitudinal or field studies to investigate whether the moral credit gained from a pro-environmental choice in one session leads to increased, less restrained use over time. 

Sixth, our focus was on the environmental impact of AI centers on energy, a topic that receives considerable attention in the literature \citep[see e.g.,][]{luccioni2024power, luccioni2024light, Li2025}. Future research should consider environmental footprints more holistically, including water consumption. Additionally, we must acknowledge the geographic inequity of AI's environmental impact and its varying severity for local societies \citep{li2024towards}. This issue is especially important in regions with limited groundwater resources, where large-scale withdrawal of groundwater by data centers could harm local ecosystems.

Finally, the growing attention to AI's energy consumption in (social) media, specifically the contrast between aggregate consumption and the marginal cost of a single query \cite[e.g.,][]{Milmo_Google_2024, ODonnel2025, kearney2025google}, may affect the effectiveness of the energy label intervention. Communication that highlights the impact of individual actions and the resource intensity of single prompts could enhance the label's effectiveness by increasing user self-efficacy. Conversely, information suggesting that an individual has negligible impact could reduce the label's effectiveness. Therefore, the intervention's success should be assessed for robustness across users with diverse perceptions of their personal impact.

\section{Conclusion}
In a recent commentary, \citet{luccioni2024light} posed the question, "\textit{Light bulbs have energy ratings — so why can’t AI chatbots?}" This study addresses this question through an experimental study. We demonstrate that adding energy labels to AI systems is, at first glance, a commendable initiative that supports the call for transparency in AI. However, it should be noted that implementing energy labels is a complex process that involves trade-offs and requires additional measures to mitigate unintended side effects. Given that millions of people use AI today, our results highlight the need for organizations to consider ECD, develop new interfaces to encourage PEB, and explore strategies to mitigate unintended consequences.

\bibliographystyle{ACM-Reference-Format}
\bibliography{00_references}

\appendix

\section{Randomization Checks}
\label{sec:appendix_randomization}
We conducted additional tests to ensure that the randomized assignment worked as intended. A chi-square test of independence was performed to determine the association between gender and  treatment, suggesting no significant difference, $\chi^2(4, \emph{N}=365) = 2.22,~\emph{p} = .695$. Moreover, we conducted a series of \emph{t}-tests between the treatment and \textit{age}, \textit{knowledge with AI}, and \textit{completion time}. The results suggest that there was no significant difference in age between groups, $\emph{t}(365) = 1.35,~\emph{p} = .179$. There was no significant difference in knowledge scores between groups, $\emph{t}(365) = 0.08,~\emph{p} = .940$. Finally, there was no significant difference in completion time between groups, $\emph{t}(365) = 0.32,~\emph{p} = .747$. Thus, we conclude an effective randomized assignment of participants. 

\section{Experimental Procedure}
\label{sec:appendix_experimental_procedure}
\begin{figure}[h]
  \centering
  \includegraphics[width=\linewidth]{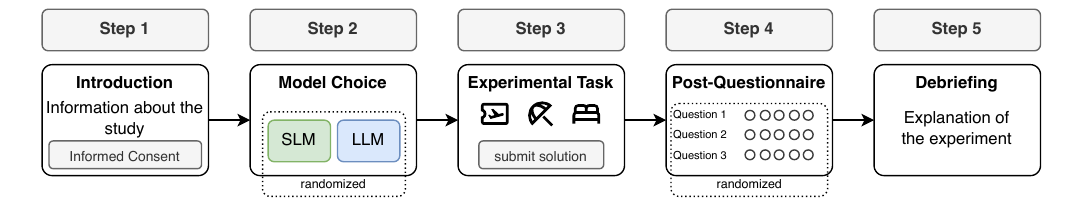}
  \caption{Overview of the Experimental Procedure}
  \Description{Overview of the Experimental Procedure}
  \label{fig:ExperimentalProcedure}
\end{figure}

The study followed a five-step experimental procedure, as illustrated in Figure \ref{fig:ExperimentalProcedure}. 

\textbf{Step 1: Introduction and Consent.} Participants were first presented with an information sheet detailing the study's objective, the approximate duration (e.g., 10 minutes), the nature of the data to be collected, and confidentiality measures. Only those who provided digital informed consent proceeded with the study.

\textbf{Step 2: Model Selection.} Participants were then randomly assigned to either the treatment or control condition. In this step, all participants were presented with an interface to select one of two AI models. They were shown the following instructions: \textit{"In the next step, you will perform a task with the help of an AI assistant. Your group has been selected to work with a new generation of AI models. Before you proceed, please read the information provided and select one of the two AI models below."}

The on-screen position (left/right) of the two models was randomized to control for order effects. A 30-second timer was enforced on this page to ensure that participants had adequate time to read the descriptions before making a choice.

\begin{figure}[h]
  \centering
  \includegraphics[width=\linewidth]{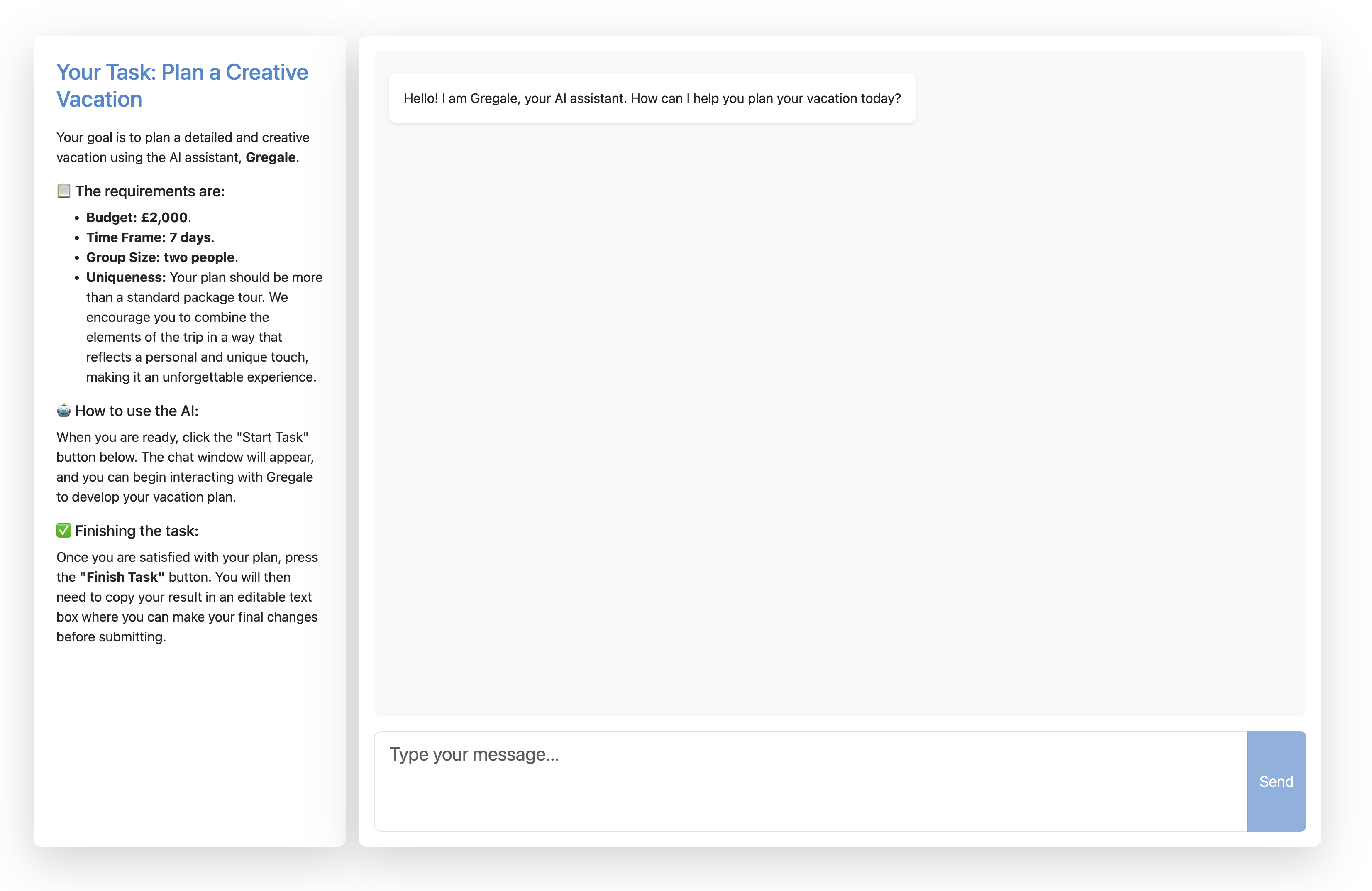}
  \caption{The main task interface. The left column displays the task instructions. The right column contains the chat interface for interacting with the AI.}
  \Description{The main task interface. The left column displays the task instructions. The right column contains the chat interface for interacting with the AI.}
  \label{fig:UITask}
\end{figure}

\textbf{Step 3: Experimental Task.} Participants were tasked with planning a 7-day vacation for two people with a budget of \textit{\pounds}2,000, using their selected AI model. We developed a task interface with a language model named GREGALE for this purpose (Figure \ref{fig:UITask}). The interface used a three-column layout: the task requirements were persistently displayed in the left column, the center column featured the chat interface for interaction with the AI, and the right column provided a text box where participants could draft, edit, and submit their final vacation plan.

\textbf{Step 4: Post-Task Questionnaire.} After submitting their plan, participants completed a post-task questionnaire to obtain quantitative information about their perception of the AI system, their attitude towards the system, and standard demographic information. The specific measures are detailed in Section \ref{sec:appendix_measurement_instrument}.

\textbf{Step 5: Debriefing.} Finally, participants were shown a debriefing page. This page thanked them for their participation, revealed the true purpose of the study, and explained the differences between the experimental conditions.

\section{Measurement Instrument}
\label{sec:appendix_measurement_instrument}
After the experiment, participants were asked to complete a survey to capture the key constructs as reported below. 
\textbf{Pro-Environmental Attitude} was measured using three items based on \cite{Shen2024}, who extracted items from the New Ecological Paradigm scale \cite{Trobe2000} and the Environmental Attitudes Inventory \cite{Milfont2010} on a 7-point Likert scale ranging from 1 = "Strongly disagree" to 7 = "Strongly agree": 1. \textit{I feel responsible to save resources whenever possible}. 2. \textit{I feel responsible to reduce energy consumption whenever possible}. 3. \textit{I feel responsible to consider the environmental effects of my work whenever possible}.

\textbf{Perceived Satisfaction} was measured using three items on a 7-point Likert scale ranging from 1 = "Very poor" to 7 = "Exceptional": 1. \textit{How satisfied were you overall with the interaction with the chatbot during the planning?}, 2. \textit{How good were the responses from the chatbot?}, and 3. \textit{How was the quality of interaction with the chatbot?} 

\textbf{Perceived Quality} was adapted from \cite{Chen2024} using a 7-point Likert scale ranging from 1 = "Strongly disagree" to 7 = "Strongly agree." The questions were as follows: 1. \textit{The responses of the chatbot were always accurate,} 2. \textit{The responses of the chatbot were always correct}, 3. \textit{The responses of the chatbot were always reliable.} 

 \textbf{Intensity of AI Usage} was measured in two ways: 1) \textit{Number of Prompts}, which is defined as the total number of prompts used by the user, and 2) the \textit{Average Tokens per Prompt} calculated as the ratio of the total number of tokens to the total number of prompts.

\textbf{Demographics} were collected based on age, gender, and highest level of education. 

\textbf{Honesty} was measured with a single item \textit{"Do you feel that you paid attention, avoided distractions, and took the survey seriously? Please answer honestly. Your answer will not affect your compensation,"} as used in \cite{Stanley2020}. Participants responded using the following options: 1. no, I was distracted, 2. no, I had trouble, 3. no, I didn't take the study seriously, 4. no, something else affected my participation negatively, or 5. yes.

A confirmatory factor analysis was conducted to ensure the reliability of the measurement instrument, including all perceptual variables. While the $\chi^2$(24) = 58.67, p = 0.0001 is significant, other practical fit indices suggest a good approximate fit between the empirical covariance matrix and the model-implied matrix. The Comparative Fit Index (CFI = .99) and the Tucker-Lewis Index (TLI = .98) both exceeded the conventional threshold of .95, indicating an excellent fit. The Root Mean Square Error of Approximation (RMSEA = .06) was also within the acceptable range. All factor loadings were significant (p < .001) and had standardized factor loadings between .87 and .94. For the subsequent multivariate analysis, we use mean values for each construct.
\end{document}